\documentclass[twoside,superscriptaddress,twocolumn,floatfix,a4paper,prl,longbibliography]{revtex4-2}
\usepackage{color}
\usepackage[]{graphicx}
\usepackage[T1]{fontenc}
\usepackage{amstext}
\usepackage{epstopdf}
\usepackage{amsfonts}
\usepackage{amsmath}
\usepackage[]{qcircuit}
\usepackage{hyperref}
\usepackage{multirow}

\newcommand\ket[1]{|#1\rangle}
\newcommand\bra[1]{\langle #1|}

\begin{document}
\title{Synergic quantum generative machine learning}

\author{Karol Bartkiewicz} \email{karol.bartkiewicz@upol.cz}
\affiliation{Institute of Spintronics and Quantum Information, Adam Mickiewicz University, 
PL-61-614 Pozna\'n, Poland}
\affiliation{Joint Laboratory of Optics of Palacký University 
and Institute of Physics of Czech Academy of Sciences, 17. listopadu 12, 
771 46 Olomouc, Czech Republic}

\author{Patrycja Tulewicz} \email{patrycja.tulewicz@amu.edu.pl}
\affiliation{Institute of Spintronics and Quantum Information, Adam Mickiewicz University, 
PL-61-614 Pozna\'n, Poland}
\affiliation{Poznan Supercomputing and Networking Center, Institute of Bioorganic Chemistry
of the Polish Academy of Sciences, PL-61-704 Poznań, Poland}

\author{Jan Roik} \email{jan.roik@upol.cz}
\affiliation{Joint Laboratory of Optics of Palacký University 
and Institute of Physics of Czech Academy of Sciences, 17. listopadu 12, 
771 46 Olomouc, Czech Republic}
\author{Karel Lemr} \email{k.lemr@upol.cz}
\affiliation{Joint Laboratory of Optics of Palacký University 
and Institute of Physics of Czech Academy of Sciences, 17. listopadu 12, 
771 46 Olomouc, Czech Republic}

\begin{abstract}
We introduce a new approach towards generative  quantum 
machine learning significantly reducing the number of hyperparameters and report on a proof-of-principle experiment 
demonstrating our approach. Our proposal depends on collaboration 
between the generators and discriminator, thus, we call it quantum 
synergic generative learning. We present numerical 
evidence that the synergic approach, in some cases, compares favorably
to recently proposed quantum generative adversarial learning.
In addition to the results obtained with quantum simulators, we also present 
experimental results obtained with an actual programmable quantum computer.
We investigate how a quantum computer implementing generative learning 
algorithm could learn the concept of a Bell state. After completing the 
learning process, the network is able both to recognize and to generate 
an entangled state. Our approach can be treated as one possible
preliminary step to understanding how the concept of quantum 
entanglement can be learned and demonstrated by a quantum computer.
\end{abstract}

\maketitle

\section{Introduction}

Generative adversarial network (GAN) machine learning is an intensely 
studied topic in the field of machine learning and artificial
intelligence research \cite{MLbook}. While quantum machine learning research is 
attracting increasingly more attention both from the industry and the 
scientific community \cite{schuld2015introduction,Biamonte2017Nature,
ciliberto2018quantum,dunjko2018machine,
PhysRevLett.122.060501,carleo2019machine,
Cai2015PRL, Chatterjee2017, Gao2018PRL, 
Rebentrost2014quantum, Schuld2019PRL, 
Bartkiewicz2019prl, McMahon614,
PhysRevLett.122.213902,shen2017deep,
Bueno:18,tacchino2019artificial,kak1995quantum,
farhi1802classification,PhysRevLett.114.140504,havlivcek2019supervised,
kandala2018extending,PhysRevLett.121.250501,PhysRevX.7.021050,
Preskill2018quantumcomputingin,PhysRevApplied.8.024030}, 
the quantum counterparts of GANs have been 
proposed in several recent papers works \cite{lloyd2018prl,
dallaire2018pra, zoufal2019npj}. For example, in the 
proposal put forward by Dallaire-Demers and Killoran in 
Ref.~\cite{dallaire2018pra}, the authors put much attention to specific
circuit ansatz and discuss methods of computing gradients in 
specific types of variational quantum circuits. It is worth noting that 
the problem of computing gradients for variational quantum circuits is 
rather complex and can be also achieved by the parameter-shift rule 
\cite{Mitarai2018,Schuld2019}. In its general form, the proposal 
of Ref.~\cite{dallaire2018pra} includes sources of entropy (i.e., bath). 

The idea behind GANs is rather simple, and it can be described with three 
circuits. The first circuit is the generator of real data $\mathcal R$, 
which is in principle an irreversible transformation depending on 
a value of a random variable $z_R$. In the case of quantum information this transformation at each instance
takes the standard input state $|0\rangle$ and outputs a labeled random 
state $\rho_\lambda.$ A good example of such a generator is a painter who 
is asked to draw a cat (the label $\lambda$ is the animal here). There 
is not a unique deterministic way of drawing a cat, nor we know how to 
construct a painter from basic elements. However, we can train a
stochastic quantum machine $\mathcal{G}$ to perform as generator 
$\mathcal{R}$ the same task by observing the output of $\mathcal{R}$ and its 
labels. However, this is not enough because $\mathcal{G}$ trained 
in this way, in general, will not be able to create new original 
instances, which can be labeled as $\lambda.$ 
Hence, an additional circuit $\mathcal{D}$ needs to be considered. 
This circuit is trained to distinguish between the samples 
$\rho_\lambda$ and the random output of $\mathcal{G},$ and it 
is referred to as \emph{discriminator.} 

The operation of $\mathcal{D}$ is optimal, if it assigns value $0$ 
to states generated by $\mathcal{R}$ and value $1$ to states generated 
by $\mathcal{G}.$ At the same time, the operation of $\mathcal{G}$ is optimal, 
if the cross-entropy between its output and states $\rho_\lambda$ is 
minimal while the discriminator is most likely to assign value $0$ to 
the output of $\mathcal{G}.$ Thus, a GAN problem is solved by adversarial
training of $\mathcal{D}$ versus $\mathcal{G}.$ The parameters of both 
the generator and discriminator can be found by numerical optimization 
or quantum gradient evaluation ~\cite{dallaire2018pra} by dividing the 
training into rounds of adversarial optimization of both generator and 
discriminator. The circuits can perform an arbitrary computation as long as they are complex enough, admitting an arbitrary 
unitary operation and measurements on a number of ancillary qubits. 
However, similarly to classical artificial neural networks, choosing 
the appropriate architectures for specific concerns is a complex problem
which is solved by trial and error. In quantum computing, this is even 
more so, because the lack of practical error correction limits the 
complexity of quantum circuits.

The quantum counterpart of the GAN (i.e., QGAN) learning similarly to its classical analogue also finds Nash 
equilibrium of two player game, where one of the players generates 
some output and the second player (discriminator, $\mathcal{D}$) tries to tell if the output is generated 
by the first player (generator $\mathcal{G}$) or provided by an external source ($\mathcal{R}$). This could be expressed 
as a min-max problem, where the statistical distance between the 
outputs of $\mathcal{G}$ and $\mathcal{R}$ is minimized over the strategies of the generator, 
whereas the distance between the outputs of $\mathcal{D}$ for $\mathcal{G}$ and $\mathcal{R}$, 
respectively, is maximized over the possible strategies of discriminator 
at the same time. In practice, this type of optimization if performed 
in rounds, and it is difficult to make the learning process stable. 
In a generative problem we do not have access directly to $\mathcal{R},$ but we can collect random samples generated by this source. However, we can formally treat it as a general multiqubit operation, where a specific unknown operation is selected according to an unknown probability distribution. 

This general approach towards QGAN employs gradient-descent methods,
as in the ansatz presented in Ref.~\cite{dallaire2018pra}.
In this standard QGAN it is impossible to apply the same sample from $\mathcal{R}$ to train both the discriminator and the generator due to the no cloning principle. Here, solve this problem by connecting the generator $\mathcal{G}$ and discriminator $\mathcal{D}$ in a single circuit. In the variational ansatz we present, we use the fact that we need to reach a conditional equilibrium state (i.e., an event when the states produced by $\mathcal{G}$ and $\mathcal{R}$ collapse on each other, yet at the same time the discriminator works at its peak performance) from the beginning of the training process.
We train such a system by increasing the probability of a circuit state collapsing to this equilibrium state. 

In this new kind of machine learning for quantum GANs, where a conceptually simpler problem is being 
solved during the training than in a typical approach to QGAN. While QGAN requires setting the hyperparmeters responsible for training the generator and the discriminator in tuns, our approach does not require this.
To introduce this approach we exploit time-reversal property of unitary transformations 
and properties of relative entropy. In particular, the approach can be understood intuitively by assuming the reversibility of the discriminator $\mathcal{D},$ which Hilbert space is the combined support space of the input state and a single-qubit decision register. 
We refer to this approach as \emph{synergic quantum generative network} (SQGEN). The reversibility condition
could be relaxed at the expense of raising the lower bound on the proposed cost function. In the extreme classical case the information on the input state is lost irreversibly in the discriminator and we cannot interpret the operation of SQGEN as conditioned on collapsing states produced by $\mathcal{G}$ and $\mathcal{R}$ on one another. Then, the cost function would be linear (instead of quadratic) in terms of the overlap between these states. This would impair the SQGEN ability to learn reproducing assemblages of density matrices instead of the mean density matrix describing the average output of $\mathcal{R}.$ In such a case, we loose the synergy between training $\mathcal{G}$ and $\mathcal{D}.$

The resulting variational quantum circuit can be trained 
using gradient methods, by means of parameter shift rules \cite{Mitarai2018,Schuld2019}
to compute partial derivatives of the cost function with respect to the circuit parameters. In many cases, it would be also practical to apply the Nelder-Mead method or similar algorithms to search for the optimal circuit parameters \cite{Jasek19}. In our experimental demonstration of SQGEN we applied the Nelder-Mead method for optimizing the circuit. For our numerical simulations of the noiseless training of larger networks we employed the BFGS algorithm.

\section{Quantum state discrimination}

\begin{figure}
\begin{flushleft}a)\end{flushleft}
\includegraphics[width=5cm]{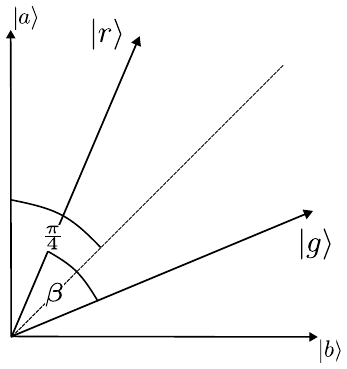}
\begin{flushleft}b)\end{flushleft}
\includegraphics[width=5cm]{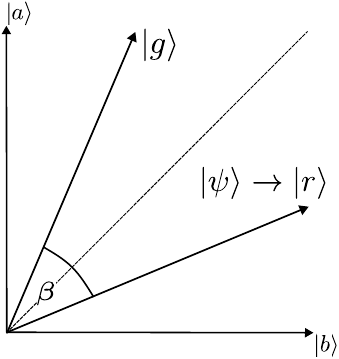}
\caption{\label{fig:1} Geometric interpretation of state-discrimination strategies. Standard measurement-based approach in basis basis $\ket{a},\ket{b}$  (a) is compared to (b) discriminator-based approach used in SQGEN. The states to be discriminated are $\ket{g}$ and $\ket{r}.$ The internal state of the discriminator associated with the optimal discrimination is denoted as $\ket{\psi}$. The discriminator is trained to find an optimal section of a Hilbert space supporting $\ket{g}$ and $\ket{r}$, where the overlap $|\langle r|\psi\rangle|^2$ is to be maximal.}
\end{figure}

The main difference between QGAN and SQGEN approaches stems from the particular strategies applied for the state discrimination~\cite{Barnett09} performed by the discriminator network, i.e., the interpretation and application of the performed measurements. 

As an introduction to state discrimination, let us assume that we want to distinguish between two states $\ket{g}$ and $\ket{r}$ containing information on the output of a generative network and the real data, respectively. These states regardless of their dimension can be represented as unit vectors on a plane. The angle between these two vectors is given as $\beta.$ The standard approach to state discrimination is finding such basis $\ket{a},\ket{b},$ where the states to be discriminated are expressed as $\ket{r}=\cos(\pi/4-\beta/2)\ket{a} + \cos(\pi/4+\beta/2)\ket{b}$ and $\ket{g}=\cos(\pi/4+\beta/2)\ket{a} + \cos(\pi/4-\beta/2)\ket{b}.$

Then, the probability of these two states being discriminated via von Neumann measurements reads $p_{a,b}= |\langle g|a\rangle|^2 |\langle r|b\rangle|^2 +  |\langle g |b\rangle|^2 |\langle r |a\rangle|^2.$ This expression can be reduced to $p_{a,b}=(1+\sin^2\beta)/2.$ This is the case QGAN training, where the optimization of discriminator consists of increasing the probability of projecting states $\ket{r}$  and $\ket{g}$ onto a state $\ket{\psi}$  co-planar with $\ket{a}$ and $\ket{b}$ while maximizing the angle $\beta$ between the discriminated states (i.e. finding the basis $\ket{a},\ket{b}$), see Fig.~\ref{fig:1}a. Sate is $\ket{\psi}$ given by a current configuration of the discriminator.

Instead of discriminating multidimensional states directly, we can introduce a single-qubit discriminator register initialized as $|0\rangle$. Now, a discriminator performs a controlled $R_y(\theta)$ on this register, where $R_y(\theta)$ is controlled by a given input of the discriminator ($|r\rangle$ or $|g\rangle$), i.e.,
\begin{equation}
R_y(\theta)\otimes|\psi\rangle\langle\psi| + \openone\otimes(\openone-|\psi\rangle\langle\psi|),
\end{equation}
where $\phi$ and $|\psi\rangle$ are parameters of the discriminator. Next, the register qubit is measured in z-basis, which yields for input $|r\rangle$ two outcomes, i.e., $-1$ with probability
$p^{(-)}_{r} =\sin^2\theta|\langle r| \psi\rangle|^2$ and $+1$ with probability 
$p^{(+)}_{r} =1-\sin^2\theta|\langle r| \psi\rangle|^2.$
The probability of optimal discrimination is given as 
$p^{(-)}_{r}p^{(+)}_{g} + p^{(-)}_g p^{(+)}_r=(1+\sin^2\beta)/2,$ if 
$|\psi\rangle=|r\rangle$ or $|\psi\rangle=|g\rangle$ and $\sin\theta=1.$ This situation is depicted in Fig.~\ref{fig:1}b.  

Both QGAN and SQGEN train the discriminator to reach its optimal performance. The advantage of SQGEN is that it automatically sets its internal pointer $\ket{\psi}$ state to $\ket{r}$, i.e., only the cases, where $\ket{r}$ collapses onto $\ket{\psi}$ and $\ket{g}$ collapses onto the support space of $\openone - |\psi\rangle\langle\psi|$ are counted as the relevant events. In case of QGAN the discriminator has to learn how to discriminate between $\ket{r}$ and $\ket{g}$ having access to only one of them at a time. This means that it performs superfluous computations that are needed for establishing a reference frame for the discrimination process. The details of the discriminator training for SQGEN together with the discriminator ansatz are discussed further in the text.

The discriminator works at its best when the probability of state discrimination is maximized. We can maximize this probability instead of the difference of rates of assigning Real/Fake label to a sample delivered by $\mathcal{R}$ or $\mathcal{G}$, as it is done in the standard GAN. This probability will be lowered, if the similarity between the samples given by $\mathcal{R}$ or $\mathcal{G}$ is increased, as it happens to the aforementioned difference of rates.

\section{Synergic quantum generative network}
 
\begin{figure}
\begin{flushleft}a)\end{flushleft}
\includegraphics[width=8cm]{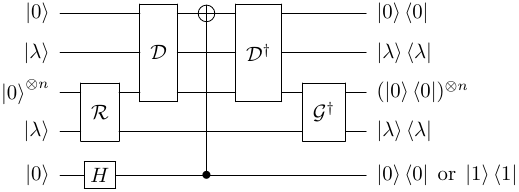}

\vspace{1mm}
\begin{flushleft}b)\end{flushleft}
\includegraphics[width=8cm]{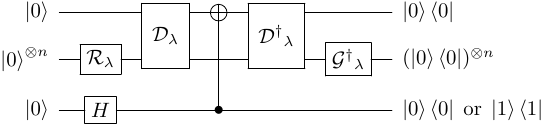}

\vspace{1mm}
\begin{flushleft}c)\end{flushleft}
\includegraphics[width=8cm]{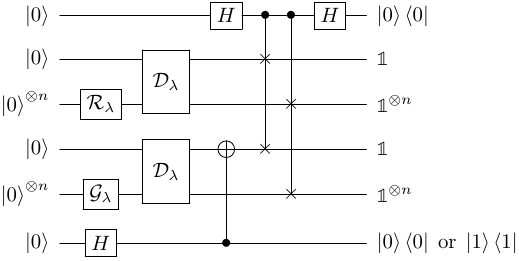}

\caption{\label{fig:2} 
In the synergic quantum generative learning protocol, 
the probability of jointly postselecting the listed states is 
proportional to the value of the cost function (\ref{eq:cost_functionJ}). 
This means that the cost function reaches its maximum value if both 
the discriminator and the generator perform their tasks optimally. 
(a) State $|\lambda\rangle$ labels the class of the output of a generator. 
It is a control state that is not changed by the operation 
of source $\mathcal{R}$ or generator $\mathcal{G}.$ For a classical label $\lambda,$ 
circuit (a) can be replaced with (b). 
In panel (c) we demonstrate an equivalent circuit inspired by a SWAP test~\cite{Barenco97}, where the measured quantity depends only on the rate of the projections of the first and the last qubits. Note that in the case 
of QGAN, in contrast to the synergic approach, one has to build (i) a circuit that compares $\mathcal{R}$ with $\mathcal{G}$, (ii) a circuit that evaluates the performance of $\mathcal{D}$ on $\mathcal{R},$
and (iii) a circuit that evaluates the performance of $\mathcal{D}$ on $\mathcal{G}.$ To compare SQGEN with QGAN we also include the bottom qubit in all the panels is measured in $Z$ basis. Depending on the outcome, we include or ignore the existence of $\mathcal{D}.$ This allows to measure either the cost function or only source-generator fidelity. No
hyperparamters are set by trial of error.
}
\end{figure}

Here, we consider reversible (unitary) 
discriminators $\mathcal D,$ which are provided with a generated state 
$\rho_{\lambda,z_R}$ its label $\lambda,$ a random variable $z_D,$ and 
a large enough ancillary Hilbert space to enable complex 
quantum computations.
The $\lambda$ parameter serves as a label for generating data states ~\cite{dallaire2018pra} and parameter $z_R$ is a random variable representing the unknown internal 
of its source, i.e., the state of generator $\mathcal R$. 
Note that $z_R$ is not accessible to the discriminator because 
the learning process must be independent of any knowledge on 
internal operation of the real generator $\mathcal R$.
The task of the discriminator is to decide, for every input, if the input 
was indeed provided by generator $ \mathcal R$ or not. 
The discriminator is trained only on a limited, but large, number 
of states $\rho_{\lambda,z_R}$ and their labels.   
Note that in the classical ML the random variable is needed
for the discriminator to make a decision if its input is 
real or fake, if the fakes are indistinguishable from
the real inputs. In the quantum case, this is not
necessary, as the collapse of a wave function of the discriminator output
will achieve the same effect.

The third component is the circuit that is the model circuit of our 
generator $\mathcal{G}$ to be trained. This generator processes the same 
type of input as generator $\mathcal R$ and is provided with an 
independent random variable $z_G.$ We denote the output of this circuit 
$\sigma = |\psi_{\lambda,z_G}\rangle\langle\psi_{\lambda,z_G}|.$ 
The action of the generator is reversible as long as we know the value 
of the random variable $z_G.$ We assume that this is the case as this is 
a classical variable.
We use random variables $z_G,z_R$ to represent the internal states of both the 
$\mathcal G$ and $\mathcal R$ generators, so we also get
random states at the output of these gates. We train the 
generator $\mathcal G$ by observing the output of the 
source $\mathcal R$, but we cannot expect the output of $\mathcal G$ 
to be perfectly correlated with $\mathcal R.$
This is because, we only minimize the relative entropy of their 
outputs, defined as 
\begin{equation}
S(\sigma_{\lambda,z_G}||\rho_{\lambda,z_R}) = \mathrm{Tr}(\sigma^2_{\lambda,z_G})
-\mathrm{Tr}(\sigma_{\lambda,z_G}\log\rho_{\lambda,z_R})
\end{equation}
or in terms of Newton–Mercator series as
\begin{eqnarray}
S(\sigma_{\lambda,z_G}||\rho_{\lambda,z_R})  &=&  \langle 1-\rho_{\lambda,z_R} \rangle 
+  \langle( 1-\rho_{\lambda,z_R})^2 \rangle/2\nonumber\\ 
&& +\langle( 1-\rho_{\lambda,z_R})^3 \rangle/3+...,
\end{eqnarray}
where $\langle \rho_{\lambda,z_R} \rangle =  \langle\psi_{\lambda,z_G}|\rho_{\lambda,z_R} |\psi_{\lambda,z_G}\rangle.$ By keeping only the first term of 
this expansion we are left with linear relative entropy $S_L$, which for 
random samples of $\sigma_{\lambda,z_G}$ and $\rho_{\lambda,z_R}$ becomes
\begin{eqnarray}
S_L(\sigma_{\lambda,z_G}||\rho_{\lambda,z_R}) =  1 - \mathrm{Tr}(\sigma_{\lambda,z_G}\rho_{\lambda,z_R}).
\label{eq:cost_function0}
\end{eqnarray}
Sample randomness (i.e., the statistics of $z_R$ and $z_G$), is required to place the linear entropy in the context of machine learning. The aim of a generative algorithm is, 
given samples $\rho_{\lambda,z_R}$ prepare samples $\sigma_{\lambda,z_G},$ which are statistically indistinguishable from new samples $\rho_{\lambda,z_R},$ not used in the training.
Thus, $S_L(\sigma_{\lambda,z_G}||\rho_{\lambda,z_R})$ should be minimized on average, i.e., over random samples denoted by $z_G$ and $z_R.$ To indicate such averaging, we drop the $z_G,z_R$ indices and from now we focus only on an average relative entropy. Note that relative entropy is in general jointly convex. In the linear approximation it is no longer the case, it is simply linear. This allows us to interpret $\rho_{\lambda}$ and $\sigma_{\lambda}$ as average density matrices of the states produced by the generators. For the generator $\mathcal G$ to mimic the source $\mathcal R$ correctly, it must also reproduce the probabilities of occurrence of the samples, not only to minimize the distance between the average states $\rho_{\lambda}$ and $\sigma_{\lambda}$. Therefore, using a discriminator is essential in our approach. While optimizing the generator $\mathcal G$, the discriminator $\mathcal D$ should reward a situation where a specific sample $\sigma_{\lambda,z_G}$ is close a single sample of $\rho_{\lambda,z_R},$ and penalize this otherwise. For this reason, the state of the discriminator must be independent of $z_G$ and $z_R.$ Moreover, assuming that a minimal achievable distance between $\rho_\lambda$ and $\sigma_\lambda$ has been reached, its cost function should be minimized if distributions of $z_G$ and $z_R$ are as similar as possible.

\subsection{Generator ansatz}
Linear entropy is directly measurable. Sometimes the second term in the expression 
is referred to as SWAP test. However, $S_L$ is alone is not enough  to correctly 
train the generator. To demonstrate this, let us consider the following example, 
where random variables $z_G,z_R$ are given via probability distributions $p$ and $q,$ 
respectively. 

Thus, the mean linear entropy, or equivalently the cost function 
of the generator reads
\begin{eqnarray}
J_G &=&  1 - \sum_{z_G,z_R} p(z_G)q(z_R)\mathrm{Tr}(\sigma_{\lambda,z_G}\rho_{\lambda,z_R})\nonumber \\
&=&1-\mathrm{Tr}(\sigma_{\lambda}\rho_{\lambda}),
\label{eq:cost_functionG}
\end{eqnarray}
where $\sigma_{\lambda}=\sum_{z_G}p(z_G)\sigma_{\lambda,z_G}$ and  
$\rho_{\lambda}=\sum_{z_R}q(z_R)\rho_{\lambda,z_R}$ are mean outputs 
of the source and the generator.

The independence of $S_L$ on $p$ and $q$ can lead to the following case. 
Assume that we have at random two states, i.e., $\rho_0 = |0\rangle\langle 0|$ 
and $\rho_1 = |1\rangle\langle 1|$ with $p(0)=p(1)=1/2$. Now, we can reach 
the same value of relative entropy by using uniform sampling either 
from $\sigma_0=|+\rangle\langle +|$ and $\sigma_1 = |-\rangle\langle -|$ 
or from $\sigma_0=|0\rangle\langle 0|$ and $\sigma_1 = |1\rangle\langle 1|.$ 
This is as expected, as the two assemblages are indistinguishable merely by measuring overlap.

\subsection{Discriminator ansatz}
To resolve between the real and fake states, we need to go beyond a simple swap test and make use of a discriminator, which would calculate the probability 
of discriminating states $\rho_{\lambda,z_R}$ and $\sigma_{\lambda,z_G}.$ 
From the standard theory of optimal state discrimination we know
that the probability of discriminating between two pure qubits can be 
expressed as $1-\cos^2(\theta(\rho_{\lambda,z_G})-\theta(\sigma_{\lambda,z_G})).$ 
This can be easily understood in terms of the Mallus law, where qubits 
are encoded as single-photon polarization. In particular, one qubit
is encoded as a linearly-polarized photon so that a polarizer can be set 
to transmit this photon. The second photon is transmitted with probability  
$\cos^2(\theta(\rho_{\lambda,z_G})-\theta(\sigma_{\lambda,z_G})).$ 
Thus, the training of the discriminator corresponds to finding 
such a function $\theta$ that the value of 
$\cos^2(\theta(\rho_{\lambda,z_G})-\theta(\sigma_{\lambda,z_G}))$ is minimized. 
This allows us to define the following cost function minimized 
by the discriminator and maximized by the generator, i.e.,    
\begin{eqnarray}
J_D &=&  1-\sum_{z_G,z_R,z_D} p(z_G)q(z_R)g(z_D)\\\nonumber
&&\times\cos^2(\theta_{z_D}(\sigma_{\lambda,z_G})-\theta_{z_D}(\rho_{\lambda,z_R})+\pi/2),
\label{eq:cost_function2}
\end{eqnarray}
where $g(z_D)$ is the probability of the discriminator having an internal state $z_D$. At the same time, we train the generator to produce an assemblage 
$\{\sigma_{\lambda,z_G},q(z_G)\}$ which maximizes 
$\mathrm{Tr}(\sigma_{\lambda,z_G}\rho_{\lambda,z_R})$
or $\cos^2(\theta_{z_D}(\rho_{\lambda,z_G})-\theta_{z_D}(\sigma_{\lambda,z_G})).$

In order to associate this function with measurable quantities, we propose the following ansatz. We work on two registers containing the state to be processed by the discriminator, i.e., an ancillary qubit initialized as $|0\rangle$ and the processed state $|\psi\rangle$. The discriminator is now described by the following unitary operator performing a $y$-axis rotation on the ancillary qubit:
\begin{equation}
D = \openone\otimes U_{z_D}|0\rangle \langle 0| U_{z_D}^\dagger  +R_y(\theta)\otimes(\openone - U_{z_D}|0\rangle \langle 0| U_{z_D}^\dagger),
\label{eq:Uzd}
\end{equation}
where $R_y(\theta)= \cos(\theta)\openone + i\sin(\theta)Y.$ Let $ U_{z_D}|0\rangle=|\phi\rangle,$ then $|\psi\rangle = \alpha |\phi\rangle + \sqrt{1-\alpha^2}|\phi_\perp\rangle,$ where $0\leq \alpha \leq 1$ and $\langle \phi |\phi_\perp\rangle=0.$
The probability of a state $|\psi\rangle$ being recognized as real by the
discriminator is given as
\begin{equation}
p(\alpha)=|\langle 0|\langle \psi | D | 0\rangle|\psi\rangle|^2 =|(1-\alpha^2) \cos\theta + \alpha^2|^2,
\end{equation}
where $p=1$ for $\alpha =1$ and arbitrary $\theta.$ 
In particular, the probability of a state $|\phi_\perp\rangle$ 
being recognized as real reads
\begin{equation}
p(0) =|\langle 0|\langle \phi_\perp | D | 0\rangle|\phi_\perp\rangle|^2 = \cos^2\theta,
\end{equation}
where $p=0$ for $\theta=\pi/2.$ 
Thus, we train the discriminator 
to have $\theta=\pi/2$ and $U_{z_D}$ which sets $|\phi\rangle \langle \phi|$
as close as possible to $\rho_{\lambda,z_G}$ (i.e., $U_{z_D}|0\rangle\langle 0|U_{z_D}^\dagger \approx \rho_{\lambda,z_G} $). From now on we will assume that  $\theta=\pi/2$ unless stated otherwise.

It can be shown by direct calculations that the expression quantifying the difference between predictions of a discriminator for two different states reads
\begin{eqnarray}
|p(\alpha)-p(\beta)| &=& |\beta^2-\alpha^2|(\beta^2+\alpha^2)\nonumber \\&=&|\sin(\theta_\alpha+\theta_\beta)\sin(\theta_\alpha-\theta_\beta)|
\end{eqnarray}
where $\cos\theta_\alpha = \alpha^2$ and $\cos\theta_\beta = \beta^2.$
This difference is maximized if either $\beta=1$ or $\alpha=1$ i.e., the discriminator is set to maximize the $p$ for a real state from assemblage  $\{\rho_{\lambda,z_R},p(z_R)\}.$ In this optimal case we arrive at the Mallus law for the discriminator, i.e.
\begin{equation}
|p(\alpha)-p(1)| = \cos^2(\theta_\alpha+\pi/2)=1-\alpha^4,
\end{equation}
where $\alpha^2 =\cos\theta_\alpha= \langle\phi_{z_D}|\sigma_{\lambda,z_G}|\phi_{z_D}\rangle$.

The optimal settings for the discriminator are provided by minimizing the distinguishability between assemblages $\{|\phi_{z_D}\rangle\langle \phi_{z_D}|,g(z_D)\}$ and $\{\rho_{\lambda,z_R},p(z_R)\},$ i.e.,
\begin{eqnarray}
J^*_D &=&  1-\sum_{z_R,z_D} p(z_R)g(z_D)\cos^2(\theta_\alpha)\nonumber\\
\label{eq:cost_function3}
\end{eqnarray}
where $\alpha^2 =\cos\theta_\alpha= \langle\phi_{z_D}|\rho_{\lambda,z_R}|\phi_{z_D}\rangle$ and $|\phi_{z_D}\rangle = U_{z_D}|0\rangle$.

If we reach the minimum of $J^*_D$ ($p=r$ and $\langle\phi_{z_R}|\rho_{\lambda,z_R}|\phi_{z_R}\rangle=1$), then for the corresponding parameters of discriminator and assemblages  $\{\rho_{\lambda,z_R},p(z_R)\}$ 
consisting of orthogonal states, we can return to the original cost function
\begin{eqnarray}
J_D &=&  1-\sum_{z_G,z_R,z_D} p(z_G)q(z_R)g(z_D)\\\nonumber
&&\times\cos^2(\theta_{z_D}(\sigma_{\lambda,z_G})-\theta_{z_D}(\rho_{\lambda,z_R})+\pi/2),
\label{eq:cost_function4}
\end{eqnarray}
where for a given assemblage $\{\sigma_{\lambda,z_g},q(z_G)\}$ at minimum of $J^*_D$ we obtain $\theta_{z_D}(\sigma_{\lambda,z_G})=\theta_\alpha$ and $\theta_{z_D}(\rho_{\lambda,z_R})=(1-\delta_{z_D,z_R})\pi/2$.
Here, $\alpha^4 =\cos^2\theta_\alpha= |\langle\phi_{z_D}|\sigma_{\lambda,z_G}|\phi_{z_D}\rangle|^2=\mathrm{Tr}(\sigma_{\lambda,z_G}\rho_{\lambda,z_D}).$
This function is now minimized over the parameters of the discriminator, regardless of the settings of the generator.

Such a discriminator is independent of the generator. However, if the
input assemblage is unknown due to the no-cloning theorem, we cannot send the real states both to the generator and the discriminator operating in parallel.
It is also impossible to train the generator and discriminator on the same set subsequently (as in traditional QGAN), as the states are destroyed during measurements. Thus, we need to design an alternative generative learning framework to QGAN.

\subsection{Synergic ansatz}

As an alternative to the standard adversarial optimization, we propose 
minimizing a single cost function, i.e.,
\begin{eqnarray}
J &=& 1-\sum_{z_G,z_R,z_D} \left[ g(z_D)p(z_G)q(z_R)\cos^2(\theta_{z_D})\right.\nonumber \\ && \left.\times \mathrm{Tr}(\sigma_{\lambda,z_G}\rho_{\lambda,z_R})\right].
\label{eq:cost_functionJa}
\end{eqnarray}
If $\theta_{z_D}=0,$ this function reduces to $J=J_G$. If $\mathrm{Tr}(\sigma_{\lambda,z_G}\rho_{\lambda,z_R})=1,$ the cost function $J$ reduces to $J^*_D.$ 
The cost function can be interpreted as probability that the assemblages $\sigma$ and $\rho$ are distinguishable for a given setting of the discriminator. This quantity is minimized if both the generator and the discriminator are optimized simultaneously. 
If we optimize only the generator or the discriminator,
there is always a place for improving $J$ by optimizing the other. 
Finally, in order to improve the readability we plot an equivalent cost function
\begin{eqnarray}
J &=& 1-2\sum_{z_G,z_R,z_D} \left[ g(z_D)p(z_G)q(z_R)\cos^2(\theta_{z_D})\right.\nonumber \\ && \left.\times \mathrm{Tr}(\sigma_{\lambda,z_G}\rho_{\lambda,z_R})\right].
\label{eq:cost_functionJ}
\end{eqnarray}

Let us again assume that the source provides at random two states, i.e., $\rho_0 = |0\rangle\langle 0|$ 
and $\rho_1 = |1\rangle\langle 1|$ with $p(0)=p(1)=1/2$. Now, if we consider two configurations of the generator corresponding to equiprobable generation ($q(0)=q(1)=1/2$) of $\sigma_0=|+\rangle\langle +|$ and $\sigma_1 = |-\rangle\langle -|$ 
or $\sigma_0=|0\rangle\langle 0|$ and $\sigma_1 = |1\rangle\langle 1|,$ we can easily verify that for some configurations of the discriminator (corresponding to its optimal operation) the latter provides a lower value of $J.$  This makes SQGEN to train the generator properly by introducing a discriminator, which is not the case when only considering generator.

\subsection{Circuit for synergic ansatz}

Let us for simplicity assume that all the probabilities $p,q,r$ correspond
to a single deterministic setting. The  probabilities $q,r$ are to be found by classical machine learning. The probability $p$ is associated with the purity of the unknown assemblage $\{\rho_{\lambda,z_R}),p(z_R)\}$. If for some $z_R$ we have $p(z_G)=1$ and $\rho_{\lambda,z_R})$ is pure, then the assemblage is pure.

Now, instead of minimizing $J$ we could equivalently maximize 
$1-J= \mathrm{Tr}(\sigma_{\lambda,z_G}\rho_{\lambda,z_R})\cos^2(\theta_\alpha).$
Such a function can be measured directly in a single circuit.
To this end, we propose connecting conjugated circuits to form
a circuit that has $\mathcal{D}$ 
interfaced with its reverse of $\mathcal{D}$ with a conditional 
$X$-gate in between (i.e., Pauli $\sigma_x$ operation) in the first 
qubit as depicted in Fig.~\ref{fig:2}a. To reduce the complexity of this 
circuit, let us note that the labels marking the class to which a given 
state belongs to can be purely classical. This means that generator 
$\mathcal{G}$ and discriminator $\mathcal{D}$ can be controlled by a 
classical variable $\lambda$, which simplifies the quantum circuit
from Fig.~\ref{fig:2}a to the one depicted in Fig.~\ref{fig:2}b.
Note that the middle (generator) qubit in Fig.~\ref{fig:2}b can in general
represent an arbitrary number of qubits, i.e., $\rho$ and $\sigma$
can be of arbitrary large Hilbert space.

The circuit in  Fig.~\ref{fig:2} with probability $\mathrm{Tr}(\sigma_{\lambda,z_G}\rho_{\lambda,z_R})$  measures $|0\rangle$ for qubits other than the first ancillary qubit. This is equivalent to projecting the fake state $\sigma_{\lambda,z_G}$ on the real state $\rho_{\lambda,z_R}=|\psi\rangle\langle\psi|.$ Thus, by postselection, we measure the following value associated with cost function $J,$ i.e., 
\begin{equation}
|\langle 0|\langle \psi | C_1 | 0\rangle|\psi\rangle|^2 =|-\sin(2\theta)(1-\alpha^2) + \alpha^2|^2=\cos^2\theta_\alpha, 
\end{equation}
where $\alpha^2 =\cos\theta_\alpha= \langle\phi_{z_D}|\rho_{\lambda,z_R}|\phi_{z_D}\rangle$ could be maximized equivalently for $\theta=\pi/2$ (discriminator regime) or $\theta=\pi/4$ (comparator regime), and $C_1=D^\dagger(X\otimes\openone) D,$
if the last qubit is projected on $|1\rangle$ or $C_0=\openone$ if the last qubit is projected on $|0\rangle.$ Then, we obtain $|\langle 0|\langle \psi | C_0 | 0\rangle|\psi\rangle|^2=1$ and we are left with a circuit
independent of the discriminator parameters. 

We have already discussed the discriminator regime $\theta=\pi/2.$ However, it is now apparent that we can also optimize the settings of the discriminator for $\theta=\pi/4.$ In such a case the probability of finding the first qubit in state $|0\rangle$ varies between $p(\alpha_{\mathrm{max}})=1$ and $p(0)=1/2$. If for a given state the discriminator outputs $p=1,$ we know that the state was recognized as originating from the source. Thus, in the comparator regime it is convenient to use a value of $p'(\alpha)=2p(\alpha)-1$ and to interpret this value as a probability of recognizing the associated state as real, as in the discriminator regime. Now, we can observe that the measured probability $p(\alpha)$ compared against the probability $p(0)$ of $|\phi_\perp\rangle$ being recognized as a real state  becomes $p(\alpha)-p(0) = p'/2,$ hence the term comparator. This difference $p(\alpha)-p(0)$ is maximized while optimizing the discriminator. Thus, it is reasonable to introduce a cost function for a discriminator which could be easily interpreted in both regimes as the probability of a given state being properly associated with its origin (i.e., $G$ or $R$), which reads  $J_{D}=1-p'(\alpha)/2.$ In the discriminator regime $p'(\alpha)=p(\alpha)$ and in the comparator regime $p'(\alpha)=2p(\alpha)-1,$
where $p$ is the measured quantity. Note that $J_{D}$ is optimized for the 
same parameters of discriminator in both regimes.

The complete circuit can be considered as working in two settings, depending on 
detecting $|0\rangle$ or $|1\rangle$ in the last qubit in Fig.~\ref{fig:2}b. In the latter case,
the linear relative-entropy between the generator and the source can be measured by feeding states 
$\rho_{\lambda,z_G}$ to the circuit and for the fixed values of $\lambda$ and 
$z_G,$ and consecutively measuring the rate at which the state of 
the generator line of the circuit is projected on $|0\rangle.$ However, 
this is only the case if the reversible discriminator returns 
$|1\rangle$ for a state generated by $\mathcal{G}$ and $|0\rangle$ for 
a state provided by $\mathcal{R}.$ The probability of this process is 
proportional to the rate at which the top line is projected onto 
$|0\rangle.$ Given that the top qubit is projected onto $|0\rangle,$ 
the middle line measures the linear cross-entropy. 
In the opposite case (the decision qubit is detected to be in $|1\rangle$), 
the operation of the discriminator failed to be reversed and 
the detection rates of the middle line are meaningless. Hence, both 
the discriminator and the generator work at their best, if the joint 
detection rates of $|0\rangle$ in both top-most circuit qubits
in Fig.~\ref{fig:2}b are maximized simultaneously. This is why we refer 
to the learning process as synergic learning. However,
there exist solutions to this optimization problem, where the generator
$\mathcal{G}$, taken separately from the discriminator, does not perform 
similarly to $\mathcal{R}.$ To address this issue, we consider the regime 
where only the similarity between $\mathcal{G}$ and $\mathcal{R}$ is 
maximized ($|0\rangle$ detected 
in the third qubit in Fig.~\ref{fig:2}b). More generally, we could 
consider the synergic learning as a process where both $\mathcal{D}$ 
and $\mathcal{G}$ are trained cooperatively, under the condition that 
$\mathcal{G}$ also is improving separately. To optimize the performance 
of the quantum setup, we propose to update 
its parameters using the Nelder-Mead algorithm or gradient descent 
to minimize the cost function (\ref{eq:cost_functionJ}). 

To consider a possible ansatz for the discriminator, let us again 
consider the regime, where the $X$ operation is active in the decision 
qubit. While maximizing the detection rates for $|0\rangle$ in the 
qubit generated state by varying the parameters of generator 
$\mathcal{G}$, we are making it less likely to detect $|0\rangle$ 
in the decision line. 
If the operations of $\mathcal{G}$ and $\mathcal{R}$ are identical, 
then gate $X$ will flip the top qubit and could not achieve maximal 
two-fold detection rates of $|0\rangle$ in both qubits, unless we allow 
$D$ to become a Hadamard gate $H$, conditioned on the similarity of 
$\mathcal{R}$ and $\mathcal{G}$ circuits. Note that, while maximizing the 
detection rates of $|0\rangle$ in the decision line by varying the 
parameters of the discriminator 
$\mathcal{D}$, in general, we do not necessarily decrease the value of 
relative entropy. If during the training the discriminator becomes 
a separable operation similar to $\sqrt{H}\otimes 1,$ and the generator 
$\mathcal{G}$ is very close to operating as $\mathcal{R}.$ 
Then, by optimizing $\mathcal{G}$ even further we would not influence 
the detection rate in the top qubit, i.e., the discriminator stops 
learning. In fact, the detection 
rate stops varying with $\mathcal{G}$ as soon as the operation 
$\mathcal{D}$ becomes separable. This suggests that inseparability 
of $\mathcal{D}$ is necessary to train the discriminator. 
Thus, it must be ensured during the design of $\mathcal{D}$ that 
its outcome in the decision qubit is strongly correlated with the 
generator qubits. This can be easily achieved by making  the discriminator 
to consist of a $Y$-rotation controlled by the generator output qubits, targeting 
only the discriminator decision qubit. This rotation is set to $\pi/2$ to
compute $p$ and $q,$ and to $\pi/4$ in case of minimizing $J.$ 
The discriminator should also admit 
arbitrary unitary transformations before the controlled operations.
This guarantees that the output of a discriminator is state-dependent, 
and the optimization works as described above.

\begin{figure*}
\includegraphics[width=16.5cm]{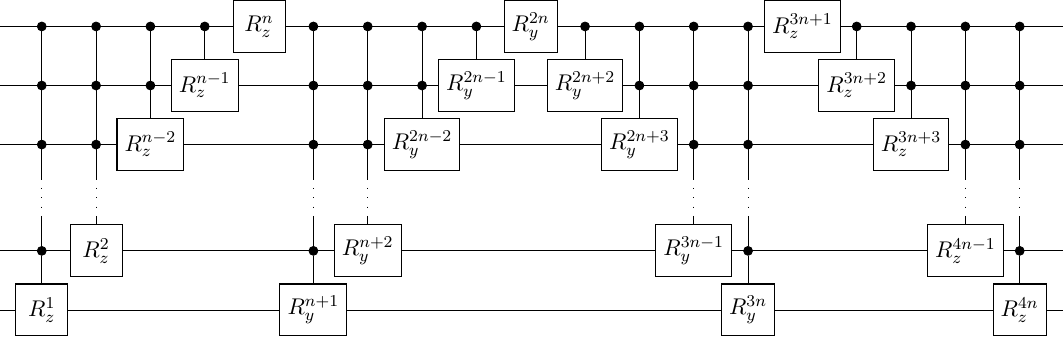}
\caption{\label{fig:3} A decomposition of an arbitrary $n$-qubit unitary gate into  $Z$ and $Y$ rotations (i.e., $R^n_z$ and $R^n_y$, respectively) controlled by multiple qubits as introduced in Ref.~\cite{Mottonen}. This circuit ansatz was used for software implementation of $\mathcal{G}$ and $\mathcal{D}$ gates for $n=1,2,3,4,5,6$. However, the final gates used in a real quantum device is obtained by automatically replacing the multiqubit controlled gates with a sequence of two- and single-qubit gates. In this ansatz, any $n$-qubit unitary gate is described by $4n$ parameters (i.e., rotation angles).
}
\end{figure*}

In our experiments and numerical simulations, we use the circuit ansatz of M\"{o}tt\"{o}nen et. al. from Ref.~\cite{Mottonen}. This means that both $\mathcal{G}$ and $\mathcal{D}$ [i.e., $U_{z_D}$ from Eq.~(\ref{eq:Uzd})] are implemented by a circuit block depicted in Fig.~\ref{fig:3}. We chose this particular ansatz because of its universality, uncomplicated implementation, and straightforward generalization to an arbitrary number of qubits. For a relatively small number of qubits, the exponential scaling in the number of CNOT gates does not constitute a problem. In higher dimensions, one can easily switch to a different ansatz, such as the so-called hardware efficient ansatz \cite{Kandala17} to avoid unfavorable scaling. In both cases, the number of parameters scales linearly with the number of qubits.

\section*{Experimental single-qubit SQGEN}

\begin{figure}
\includegraphics[width=8.5cm]{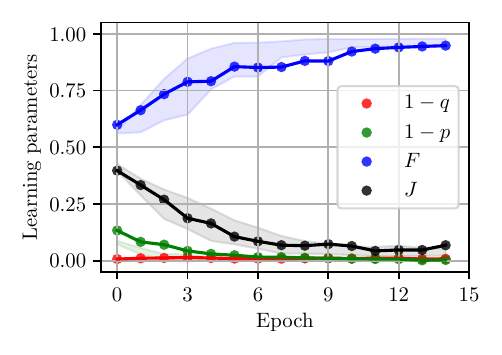}
\caption{\label{fig:4} 
The shaded areas depict the range of values obtained in 100 Monte Carlo simulations accounting for both shot noise and transmon decoherence for the quantum processor \textit{ibmq{\_}montreal}~\cite{IBMQ} using three qubits. The connected points represent the actual values measured in the training process performed as described in the main text. For the $J$ cost function, the region associated with the noise model does not include all the experimental data. This means that the noise model provided for the quantum processor by the manufacturer may be inadequate for circuits of depth of order 27. This is similar for the
source-generator $F$ and probability $p$ ($q$) of the discriminator recognizing the source (generator) state as real.
}
\end{figure}
Let us consider a proof-of-principle experiment, where $\lambda$ labels 
the bases in which states are prepared. If $\lambda=x$ the generator 
prepares $\mathcal{R}$ at random state $(|0\rangle+|1\rangle)/\sqrt{2}$ or 
$(|0\rangle-|1\rangle)/\sqrt{2}.$ The eigenstates of the remaining Pauli matrices 
$\sigma_y$ and $\sigma_z$ are prepared if $\lambda=y,z.$ This in general requires feeding 
generators $\mathcal{R}$ and $\mathcal{G}$ with uncorrelated bivariate 
random variables $z_R$ and $z_G$ (baths), respectively. In addition, 
we require that the SQGEN performs equally well for all combinations 
of values of the random variables. Let us train a SQGEN with $\mathcal{R}$ 
set as a Hadamard matrix proceeded by $X^{z_G}$ operation, i.e., 
$\lambda=x$. To make the training process more transparent, let us
focus on the special case of $z_g=0,$ only $(|0\rangle+|1\rangle)/\sqrt{2}$
is generated by $\mathcal{G}$.

In the experiment, we deal with finite numbers of shots, which 
can lead to random fluctuations in the measured values of the minimized 
cost function. To establish 
a sufficient number of shots, we analyzed the impact of this Poissonian 
noise on the experimental data. 
In the case considered, we used the Nelder-Mead algorithm because in the noise experiment, it gives better results than the gradient method, needing fewer steps to find the solution.
From our numerical simulations, it follows that for our specific problem the training to perform well 
already for about $10^4$ shots for about $10^2$ evaluations 
of the cost function. When using more than $10^6$ shots 
the performance of Nelder-Mead algorithm further improves, reaching $70$ 
cost function evaluations needed to find the minimum of the cost function. 
The speed of the convergence of this algorithm for this particular 
problem can be slightly improved by choosing a larger initial simplex. 
The requirements on the number of function evaluations and the number 
of coincidences make it feasible to implement conjugated SQGEN on 
a contemporary quantum computer. The results of the experiment are shown in Fig.~\ref{fig:4}.

We performed our experiments on \textit{ibmq{\_}montreal} quantum processor \cite{IBMQ}. 
Note that due to technical solutions used in IBMQ processors \cite{IBMQ} we cannot 
directly implement the circuit given in Fig.~\ref{fig:2}b. 
The processors, physically implement controlled-phase gates, 
controlled-not gates, and single-qubit rotations. This results in a 
circuit that performs 27 steps (circuit depth 27, 3 qubit circuit) before 
evaluating the cost function $J$. Independent 3 experiments were used to 
measure 16 values of real/fake state fidelity $F$ (circuit depth 15, 1 
qubit circuit), probability $p$ of a real state  (generated by 
$\mathcal{R}$) being classified by $\mathcal{D}$ as being real (circuit 
depth 11, 2 qubit circuit), probability $q$ of a fake state (generated by 
$\mathcal{G}$) being classified as being real (circuit depth 11, 2 qubit 
circuit). These experiments were performed for parameter values found 
after each epoch of training.

For 32000 shots such circuit runs for 15~s per single cost function
evaluation. For the random starting point used in Fig.~\ref{fig:4}, on 
average, we need 260 evaluations of the cost function to
complete 15 training epochs (an epoch corresponds to 5 iterations of the 
Nelder-Mead algorithm). Our results show that the SQGEN training on a 
quantum processor (see Fig.~\ref{fig:4}) performs similarly as predicted 
by our numerical simulations. We did not use gradient-based approach here, 
as our experience shows that it is lest robust to experimental noise and 
because of this its convergence in many cases is worse than the Nelder-Mead methods.

The experimental results, shown in Fig.~\ref{fig:4}, demonstrate 
that SQGEN can can be implemented using the available quantum computers, 
even without applying error correction. However, to obtain our result we 
applied standard measurement error mitigation, a method which corresponds to 
calibrating the detection part of the quantum computer.

To find the smallest number of shots needed for the learning process to complete, 
we have tested the proposed algorithm both on real quantum processors
and simulators available to researchers via the IBMQ project \cite{IBMQ}. Each evaluation of the
circuit was performed on 8192 shots, which was found to be sufficient to
limit the effect of Poisson noise. Due to the technical imperfections
of these real devices, the algorithm converged only in about one half
of the runs. It should be stressed out, however, that the user can always
rerun the algorithm until it converges. 
One can observe that the algorithm converges to a
non-zero value of the object function, which we also attribute to the
experimental noise in the processor. Note that using the noiseless simulators,
the algorithm converged on every attempt and the final object function
was minimized below 0.001. This supports our finding that the algorithm
is performing well, and the convergence difficulties are solely due to
the noise in real presently available quantum processors.

\section{Comparison of QGAN and SQGEN: Generating and recognizing 
a multiqubit entangled state}

The proposed approach to generative quantum learning is
conceptually different from the approaches described in 
Refs.~\cite{dallaire2018pra,zoufal2019npj}.
Both approaches can solve an interesting problem, i.e., given samples
of an entangled state, they can learn to generate the entangled states 
on their own.
Moreover, the respective discriminators can be trained to detect the entangled state. 
However, from our numerical simulations it follows that for the same 
number of cost function calls, it is the SQGEN that will complete the training first. 

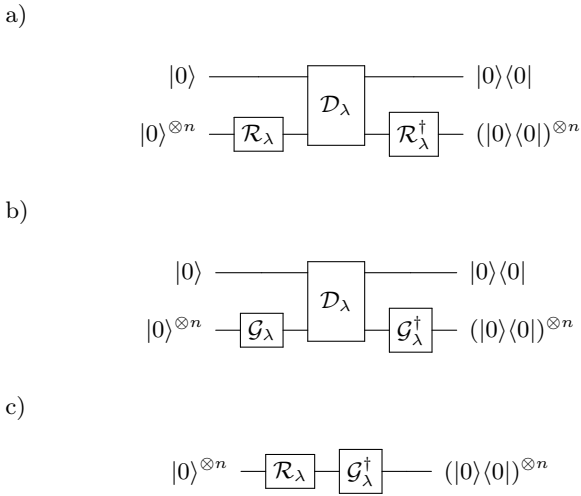
\begin{figure}
\begin{flushleft}a)\end{flushleft}
\quad\!\!\!\!\!\!\!
\Qcircuit @C=1em @R=1em {
\lstick{\ket{0}} 
& \qw 
& \multigate{1}{\mathcal{D}_\lambda} 
& \qw 
& \rstick{\ket{0}\bra{0}}\qw\\
\lstick{\ket{0}^{\otimes n}} 
& \gate{\mathcal{R}_\lambda}
& \ghost{\mathcal{D}_\lambda}    
& \gate{\mathcal{R}^\dagger_\lambda}
& \rstick{(\ket{0}\bra{0})^{\otimes n}}\qw \\
}

\vspace{2mm}
\begin{flushleft}b)\end{flushleft}
\quad\!\!\!\!\!\!\!
\Qcircuit @C=1em @R=1em {
\lstick{\ket{0}} 
& \qw 
& \multigate{1}{\mathcal{D}_\lambda} 
& \qw 
& \rstick{\ket{0}\bra{0}}\qw\\
\lstick{\ket{0}^{\otimes n}} 
& \gate{\mathcal{G}_\lambda}
& \ghost{\mathcal{D}_\lambda}    
& \gate{\mathcal{G}^\dagger_\lambda}
& \rstick{(\ket{0}\bra{0}) ^{\otimes n}}\qw \\
}

\vspace{2mm}
\begin{flushleft}c)\end{flushleft}
\quad\!\!\!\!\!\!\!
\Qcircuit @C=1em @R=1em {
\lstick{\ket{0}^{\otimes n}} 
& \gate{\mathcal{R}_\lambda}
& \gate{\mathcal{G}^\dagger_\lambda}    
& \qw 
& \rstick{(\ket{0}\bra{0})^{\otimes n}}\qw\\
}
\caption{
\label{fig:5} 
The circuits used for QGAN.
a) a circuit that evaluates the performance of $\mathcal{D}$ on $\mathcal{R}$ (computes $p$), 
b) a circuit that evaluates the performance of $\mathcal{D}$ on $\mathcal{G}$ (computes $q$),
and c) a circuit that compares $\mathcal{R}$ with $\mathcal{G}$ (computes $F$).
These circuits are used in rounds where for a given number of steps either $\mathcal{D}$ or $\mathcal{G}$ is optimized while keeping
the parameters of the other fixed. A proper setting of this procedure requires a trial of error.}
\end{figure}

To illustrate this, let us consider generator 
$\mathcal{R},$ which prepares a maximally entangled (for $n>1$) $n$-qubit GHZ state $|\Psi\rangle=(\ket{0}^{\otimes n}+\ket{1}^{\otimes n})/\sqrt{2}.$ Thus, there is one possible value of 
$\lambda=e$. The goal of the QGAN and SQGEN training is to train 
generator $\mathcal{G}$ (i.e., find the optimal circuit parameters) 
without knowing the algorithm used by $\mathcal{R}$ nor its internal 
state $z_R$ by optimization of both the discriminator $\mathcal{D}$ and the generator $\mathcal{G}.$  The circuits used for QGAN and SQGEN are shown in Fig.~\ref{fig:5}a-c and Fig.~\ref{fig:2}b, respectively. 

To compare the dynamics of the training process, we use (as in the 
previous section) three figures of merit:
(i) probability $q$ of the fake ($\mathcal{G}$-generated) state to be 
recognized as real by discriminator $\mathcal{D},$ (ii) probability $p$ 
of the real state ($\mathcal{R}$-generated) state to be recognized as 
real by discriminator $\mathcal{D},$ (iii) the distance between $D=1-F$ 
the $\mathcal{G}$-generated and $\mathcal{R}$-generated states (linear 
entropy).

Here for noiseless numerical simulations we use a gradient descent method (i.e., BFGS), which guarantees at most as many function evaluations as the Nelder-Mead method. 
The learning process for both SQGEN and QGAN is performed for a fixed number of epochs.
Each training epoch for SQGEN corresponds to a single iteration of BFGS algorithm used to minimize the cost function $J$. The relative number of iterations in QGAN is a hyperparameter that we tuned by trial of error. In the case of QGAN each epoch corresponds to one iteration of BFGS used to train the discriminator (to maximize a cost function proportional to $|p-q|$ and $p+q$) and a single iteration of BFGS to minimize $F.$

\begin{figure*}
\begin{flushleft}a)\end{flushleft}
\vspace{-0.5cm}
\includegraphics[width=6.5cm]{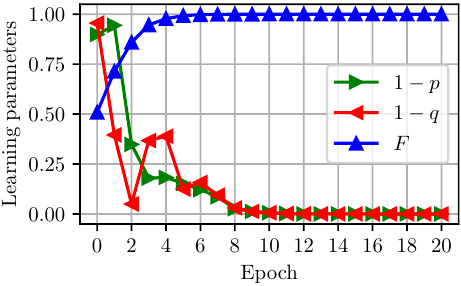}\hspace{1cm}
\includegraphics[width=6.5cm]{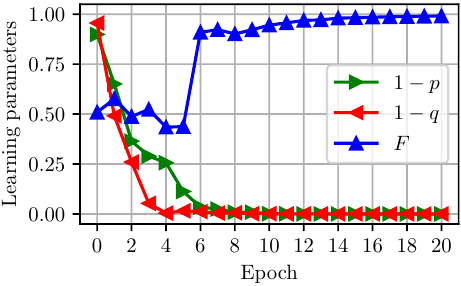}
\begin{flushleft}b)\end{flushleft}
\vspace{-0.5cm}
\includegraphics[width=6.5cm]{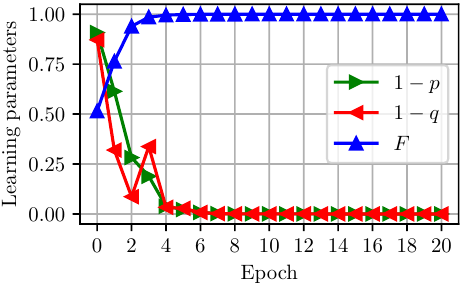}\hspace{1cm}
\includegraphics[width=6.5cm]{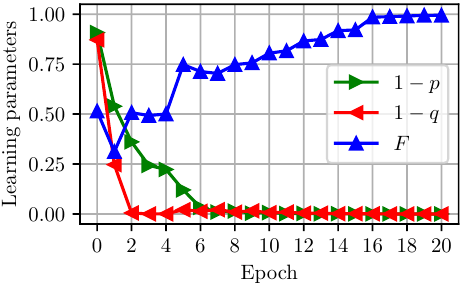}
\begin{flushleft}c)\end{flushleft}
\vspace{-0.5cm}
\includegraphics[width=6.5cm]{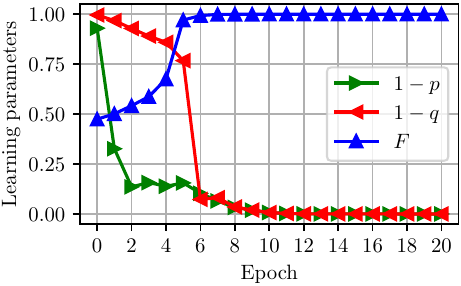}\hspace{1cm}
\includegraphics[width=6.5cm]{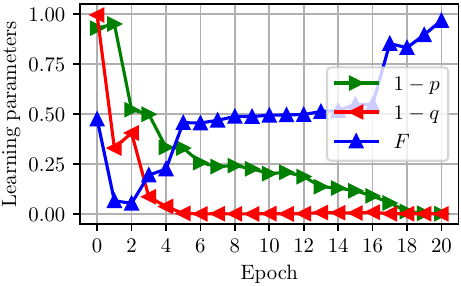}
\begin{flushleft}d)\end{flushleft}
\vspace{-0.5cm}
\includegraphics[width=6.5cm]{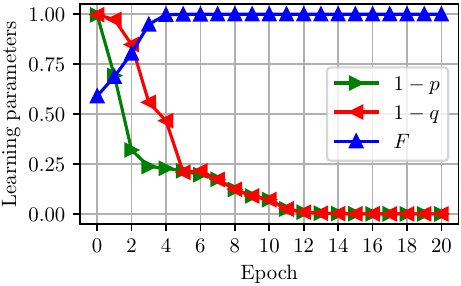}\hspace{1cm}
\includegraphics[width=6.5cm]{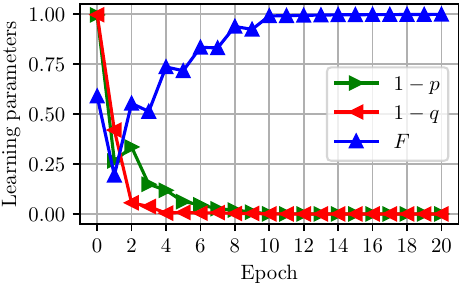}
\begin{flushleft}e)\end{flushleft}
\vspace{-0.5cm}
\includegraphics[width=6.5cm]{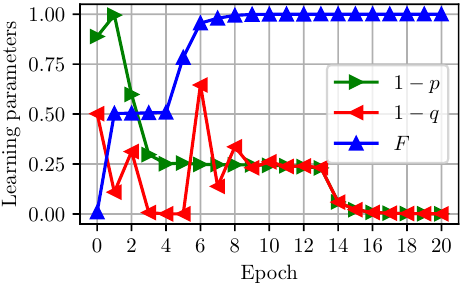}\hspace{1cm}
\includegraphics[width=6.5cm]{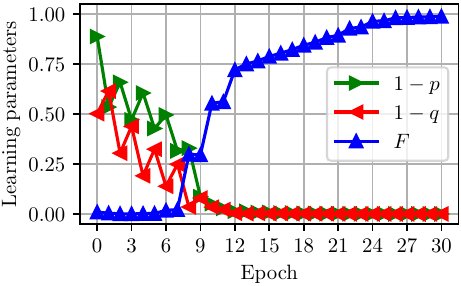}

\caption{\label{fig:6} 
Comparison of the dynamics of the  QGAN (left column) and SQGEN (right column) training process for the source providing $n$-qubit GHZ state. The sequence of panels corresponds to $n,$ i.e., (a) $n=2$, (b) $n=3$, (c) $n=4,$ etc.
We do not plot $J$ here, as QGAN has no counterpart of it. The plots illustrate the probability $p$
of a source state being recognized as real by the discriminator, probability $q$ of a generated state being recognized, and fidelity $F$ comparing the trained generator and the source.
For $n=6,$ there we can observe QGAN almost settling at a suboptimal solution. This was circumvented by not only minimizing $1-|p-q|$ in the discriminator training phase, but also by including a term proportional to $1-p$ in the cost function. No such manipulation was required for SQGEN, but including a term proportional to $1-F$ in the cost function is possible as it could reduce the number of epochs needed for the convergence of SQGEN algorithm. 
}
\end{figure*}

For SQGEN, $n+1$ qubits are required to solve the problem ($n+2$
 to also monitor $F$ in addition to $J$), while for QGAN this value corresponds to as much as $3n+2$ qubits (this includes two sources $\mathcal{R}$). Even in the SWAP test based circuit, SQGEN requires fewer qubits than QGAN,  i.e., $2n+4$ when monitoring $F.$

Our numerical investigations suggest that it takes fewer epochs for SQGEN to reach a stable probability values, and QGAN approach is slightly faster to settle on a high fidelity values. If the number of parameters is not too large and the hyperparameters of QGAN are set in an optimal way, the overall performance of SQGEN and QGAN is similar. 

However, for GHZ states, QGAN appears to reach the optimal solution for a larger number of initial setup configurations and for a fixed number of training epochs. It is hard to state this with certainty due to a limited number of tested initial configurations. For each studied value of $n$ we found optimal generator and discriminator configurations using both approaches. For each $n>1$ we also found cases where either QGAN or SQGEN settled at suboptimal values.

In Fig.~\ref{fig:6}, showing the dynamics of the learning 
process optimized after a set of function calls corresponding to 
epochs, we see the comparison of QGAN and SQGEN results for the best achieved configuration 
of QGAN hyperparameters.  More details on these simulations are summarized in Tab.~\ref{tab:1}.

\begin{table}[]
\caption{\label{tab:1} Comparison of performance of QGAN and SQGEN for 20 epochs of learning with BFGS optimizer for varied size of generated $n$-qubit GHZ state. The total run time is given in seconds, and it may vary depending both on software and hardware. The run times here were obtained as averages over 5 runs (for various initial configurations) on a workstation equipped with Intel(R) Xeon(R) CPU X5690  @ 3.47GHz, using Python-based programs utilizing, e.g., qiskit, numpy, and scipy modules. The tabulated data corresponds to Fig.~\ref{fig:6}}
\begin{tabular}{|l|l|c|cc|}\hline
\multirow{2}{*}{$n$} & \multirow{2}{*}{Feature} & \multirow{2}{*}{SQGEN} & \multicolumn{2}{c|}{QGAN} \\ \cline{4-5}
                  &                   &                   &     \multicolumn{1}{c|}{Discriminator}      &     Generator   \\ \hline
                 \multirow{3}{*}{1} &        Experiments per epoch           &    32.68               &    \multicolumn{1}{c|}{19.3}         &    130.73      \\ \cline{2-5}
                  &   Circuit depth                &         27          &   \multicolumn{1}{c|}{19}        & 5         \\ \cline{2-5}
                  &   Time per epoch                &   0.93                &     \multicolumn{2}{c|}{1.52}               \\ \hline
                 \multirow{3}{*}{2} &        Experiments per epoch           &                  61.01 &    \multicolumn{1}{c|}{46.8}         &  124.86        \\ \cline{2-5}
                  &   Circuit depth                &     92              &   \multicolumn{1}{c|}{71}        &  18         \\ \cline{2-5}
                  &   Time per epoch                &    10.87               &     \multicolumn{2}{c|}{9.97}               \\ \hline
                 \multirow{3}{*}{3} &        Experiments per epoch           &  92                  &    \multicolumn{1}{c|}{70.72}         &  147.25         \\ \cline{2-5}
                  &   Circuit depth                &     317              &   \multicolumn{1}{c|}{239}        &  55        \\ \cline{2-5}
                  &   Time per epoch                &  52.87                 &     \multicolumn{2}{c|}{39.80}               \\ \hline
                 \multirow{3}{*}{4} &        Experiments per epoch           & 136.42                   &    \multicolumn{1}{c|}{102}         &  229.19         \\ \cline{2-5}
                  &   Circuit depth                &      976             &   \multicolumn{1}{c|}{747}        &  174        \\ \cline{2-5}
                  &   Time per epoch                &      229.44             &     \multicolumn{2}{c|}{174.34}              \\ \hline
                 \multirow{3}{*}{5} &        Experiments per epoch           & 127.69                   &    \multicolumn{1}{c|}{156.66}         &  172.83         \\ \cline{2-5}
                  &   Circuit depth                &    2404               &   \multicolumn{1}{c|}{1851}        &  434        \\ \cline{2-5}
                  &   Time per epoch                &     516.45              &     \multicolumn{2}{c|}{515.85}               \\ \hline
                 \multirow{3}{*}{6} &        Experiments per epoch           & 176.79                   &    \multicolumn{1}{c|}{137.5}         &    158.44      \\ \cline{2-5}
                  &   Circuit depth                &  5385                 &   \multicolumn{1}{c|}{4159}        &  979         \\ \cline{2-5}
                  &   Time per epoch                &     1558.58              &     \multicolumn{2}{c|}{999.93}              \\ \hline

\end{tabular}
\end{table}

\section*{Conclusions}
We have proposed a new, efficient approach towards generative quantum machine learning. We have tested the proposed SQGEN algorithm experimentally on
a small-scale programmable quantum processor. The experimental results shown in Fig.~\ref{fig:4} confirm the feasibility of implementing SQGEN on a NISQ device. We have also performed feasibility study for larger experiments. However, we observed that experimental noise for $n>1$ prohibited reaching the convergence of the optimization procedure within the observed number of training epochs. 

In addition to being conceptually different from a QGAN, SQGEN in all the cases investigated numerically required fewer cost function evaluations (experiments) per training epoch than QGAN. Note that SQGEN computes only $J$ and  $p,q,F$ are computed additionally at the end of each epoch to facilitate our comparison with QGAN. Running a stable QGAN optimization is hard, as one has to carefully tune number of rounds and other parameters for training the discriminator and generator. Thus, the computational overhead of QGAN is in practice even greater. However, when properly tuned QGAN
can demonstrate some advantage over SQGEN depending on a problem dimensionality and the initial choice of circuit parameters. Both methods sometimes settle
at suboptimal solutions. The proposed SQGEN might be a good choice, if we do not want to deal with finding adequate values of many hyperparameters. Finally, SQGEN in contrast to QGAN does not require two copies of $\mathcal{R},$
which is important due to the no-cloning principle.

Note that in our numerical simulations, we have investigated how a quantum computer could learn the concept of a GHZ state. After the training, the network is able both to recognize and to generate this state. A next interesting step in would be to extend the notion of GHZ state to an arbitrary entangled state to investigate how the concept of entanglement 
could be learned and understood by a quantum computer. Solving this problem would require combining the presented concepts and methods with possibly more sophisticated classical machine learning to deal with providing labels for multidimensional, multiparty entanglement.

At this point, it is also important to stress that SQGEN cannot be directly reduced to a simple SWAP test, which corresponds to measuring only linear entropy. A SWAP test has an advantage when we are dealing with a source delivering a single pure state, but for general assemblages it not be sufficient to properly train the generator [see the discussion below, Eq.~(6)]. However, the depth of the proposed SQGEN circuit can be potentially reduced (depending on the particular circuit ansatz) by applying the circuit depicted in Fig.~\ref{fig:2}c. The multi-qubit controlled-SWAP gate can be composed of $n$ standard controlled-SWAP gates (i.e., Fredkin gates). This itself adds to the total circuit depth and at the same time increases exponentially the Hilbert space, which makes it difficult to simulate such circuits. However, using the SWAP test approach can reduce the time needed to evaluate $J$ on a quantum computer with respect to the sequential circuit studied here, but by no more than a half. The SWAP test approach to SQGEN could apply
to mitigate to some extent dissipation in real quantum devices by reducing the impact of decoherence, which accumulates over time.

To some extent, we can compare the operation of the SQGEN circuit presented in this paper to that of an uncompressed autoencoder. Just as in the case of the autoencoder, the encoder it is trained together with the decoder, so in the circuit discussed here we train the generator and discriminator together. Another common aspect is the optimization of state fidelity between the input and the output. The key difference between the autoencoder and the problem at hand, is that due to the negation gate appearing, the right side of the circuit (regarded as a decoder) shown in Fig.~\ref{fig:2}, cannot be interpreted as the inverse of its left side (treated approximately as an encoder).

The aim of a generative algorithm is to generate samples that fit the properties of the real samples without knowing the ground truth (probability density function) about how the real samples are prepared. This is not the same as memorizing the real samples and generating them. We demonstrate that our approach is able to reproduce the real quantum samples and to distinguish between similar and dissimilar samples. In our analysis the ground truth about how the real samples were prepared was relatively simple. Thus, we were able to demonstrate that SQGEN works as intended. Demonstrating that SQGEN can handle more complex data patterns requires additional research as is beyond the scope of this paper. 

Finally, it is interesting to consider some analogies between SQGEN and kernel-based machine learning. The initial part of the circuit can be viewed as the state preparation step, whereas the second part (including the $X$ gate) can be interpreted as kernel evaluation circuit. Now, the SQGEN ansatz, as in the case of kernel-based methods, can be understood as a procedure consisting of measuring Gram matrix elements. However, the main difference is that contrary to standard kernel-bases approaches, we are not interested in evaluating Garm matrix elements for a specific fixed feature map and specific pairs of points in the feature space. In our case, the kernel is generated by both parameters of the source state (associated with the generator circuit) and the parameters of the discriminator. The circuits parameters are variables that we optimize and not fixed points in the feature space. The points are given by the generator and the source. Thus, in the variational circuit, we search for such a kernel that minimizes $J$ with respect to circuit parameters. However, the circuit parameters appear with some weights which must be found by a classical algorithm, as in the case of standard applications of kernel methods.Thus, the SQGEN circuit could be considered as a generative kernel learning method which are being currently studied as a promising tool
for generative learning \cite{10.1016/j.neucom.2022.02.053}.

\section*{Acknowledgments} 
We would like to thank Dawid Maskalaniec and Mateusz Slysz 
for stimulating discussions and for their contributions
to numerical calculations.
Authors acknowledge financial
support by the Czech Science Foundation under
the project No. 19-19002S. The authors
also acknowledge the project No.
CZ.02.1.01./0.0/0.0/16\textunderscore 019/0000754 
of the Ministry of Education, Youth and Sports of the Czech Republic 
financing the infrastructure of their workplace. 
P.T. is supported by the Polish National Science
Centre (NCN) under the Maestro Grant No. DEC-2019/34/A/ST2/00081.
J.R. acknowledges the internal Palacky University grant DSGC-2021-0026.
We acknowledge the use of IBM Quantum services for this work. The views expressed are those of the authors, and do not reflect the official policy or position of IBM or the IBM Quantum team.

\section*{Appendix} 

\section*{Example: Minimal circuit}
\begin{figure}
\vspace*{2mm}
\begin{flushleft}a)\end{flushleft}
\quad
\Qcircuit @C=1em @R=1.1em {
& 
& 
&\mbox{$\mathcal{D}_\lambda$} 
& 
&\mbox{$\mathcal{D}^\dagger_\lambda$}& \\
&\qw 
&\gate{A(\lambda)} 
&\gate{U(\lambda)}  
&\gate{X} 
&\gate{U^\dagger(\lambda)} 
&\gate{A^\dagger(\lambda)} 
&\qw 
&\qw \\
&\gate{\mathcal{R}} 
&\gate{B(\lambda)} 
&\ctrl{-1}
& \qw 
& \ctrl{-1}
& \gate{B^\dagger(\lambda)} 
&\gate{\mathcal{G}_\lambda^\dagger} 
& \qw \gategroup{2}{3}{3}{4}{1em}{--} \gategroup{2}{6}{3}{7}{0.7em}{--}
}
\vspace*{1mm}

\begin{flushleft}b)\end{flushleft}
\quad
\Qcircuit @C=1em @R=1.1em {
&\qw 
&\gate{A(\lambda)} 
&\gate{X U^\dagger(\lambda)XU(\lambda)}  
&\gate{X}  
&\gate{A^\dagger(\lambda)} 
&\qw 
&\qw \\
&\gate{\mathcal{R}} 
&\gate{B(\lambda)} 
&\ctrl{-1} 
&\qw  
&\gate{B^\dagger(\lambda)} 
&\gate{\mathcal{G}^\dagger_\lambda} 
&\qw
}
\caption{\label{fig:7} 
Two equivalent circuits for training a SQGEN for single-qubits, for active $X$ gates. To reduce the total number of controlled gates, we propose not to use the third qubit as in Fig.~\ref{fig:1}, but rather to implement $\mathcal{R}\mathcal{G}^\dagger$ transformation to monitor the performance of the generator $\mathcal{G}$ by tracing out the first qubit and analyzing only the counts for the second qubit. In general, this is possible for certain discriminators. In circuit (b) the number of the controlled operations is reduced in comparison to circuit (a) by applying the reasoning similar to the one from Ref~\cite{Lemr2015prl}.}
\end{figure}
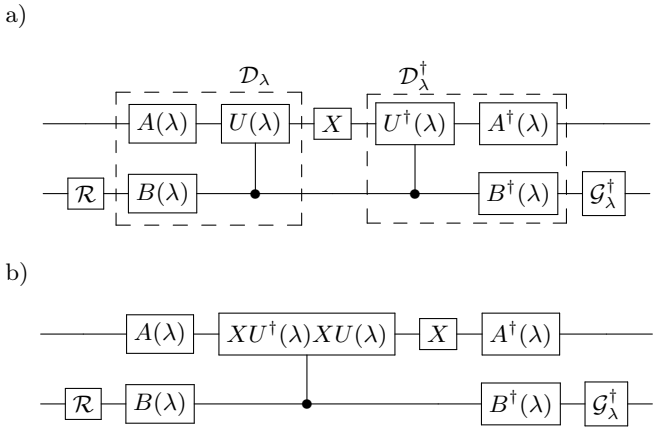

Now, let us apply the general circuit from Fig.~\ref{fig:1}b to the simplest 
case, where each circuit mode corresponds to a single qubit. In this 
case the generator $\mathcal{G}_\lambda$ can be an arbitrary 
single-qubit unitary matrix. Similarly, in this regime, the generator 
is a general two-qubit unitary gate~\cite{Lemr2015prl}. Two equivalent 
circuits implementing this operation are depicted in Fig.~\ref{fig:7}. 

When designing a dedicated quantum circuit it could be beneficial to
to replace the $XU^\dagger XU$ operation from Fig.~\ref{fig:7} with 
a controlled-phase gate implementing a controlled rotation 
$R(\hat{z},\beta)$ around $z$-axis  and single-qubit rotation 
$R(\hat{a},\alpha)$ around an axis given by versor $\hat{a}$. 
Then, in order to train QGAN it is necessary to find parameters for 
which 
\begin{equation}
XU^\dagger XU = e^{i\omega} R(\hat{a},\alpha)R(\hat{z},\beta)
R^\dagger(\hat{a},\alpha).
\end{equation}
Any unitary operation can be expressed as a rotation and a phase shift, 
i.e., 
\begin{equation}
XU^\dagger XU = e^{i\omega}\left(\cos\frac{\delta}{2} 
-i \hat{n}\cdot\vec{\sigma}\sin\frac{\delta}{2}\right)
=e^{i\omega}R(\hat{n},\delta),
\end{equation}
where
\begin{eqnarray}
-2i e^{i\omega} n_j\sin{\frac{\delta}{2}}&=& 
\mathrm{Tr}(XU^\dagger XU \sigma_j),\nonumber\\
2 e^{i\omega}\cos\frac{\delta}{2}  &=& 
\mathrm{Tr}(XU^\dagger XU),\label{eq:solve}
\end{eqnarray}
and $\sigma_j$ for $j=x,y,z$ are the Pauli matrices.
These equations allow to express the parameters of  matrix $U$ in terms 
of $\omega,\hat{n},$ and $\delta.$  Moreover, from the cyclic property 
of the trace we learn that $n_x=0$ and $\omega=0.$

Next, let us write 
\begin{equation}
R(\hat{n},\delta)=R(\hat{z},\gamma)R(\hat{y},\eta)R(\hat{z},\delta) 
R^\dagger(\hat{y},\eta)R^\dagger(\hat{z},\gamma),
\end{equation}
by using Pauli matrix algebra we can show that
\begin{equation}
n_x = \sin\eta\cos\gamma,\quad n_y = \sin\eta\sin\gamma,\quad 
n_z = \cos\eta,
\end{equation}
which allows us to compute $\eta$ and $\gamma$ for any $\hat{n}.$
Having $n_x=0,$ we get $\gamma=\pi/2.$ We also conclude that 
$R(\hat{a},\alpha)=R(\hat{z},\pi/2)R(\hat{y},\eta)$ and $\beta=\delta.$ 
Similarly, we can express the remaining operators, i.e.,
\begin{equation}
A = R(\hat{n}_a,\delta_a),\quad B = R(\hat{n}_b,\delta_b),\quad 
\mathcal{G} = R(\hat{n}_g,\delta_g),
\end{equation}
and
\begin{equation}
n_{l,x} = \sin\eta_l\cos\gamma_l,\, n_{l,y} = \sin\eta_l\sin\gamma_l,
\, n_{l,z} = \cos\eta_l,
\end{equation}
for $l=a,b,g.$ Now, for a given $\lambda,$ we can select the training 
parameters of the discriminator to be 
$\vec{\theta}_{\mathcal{D},\lambda}=
(\delta,\eta,\delta_a,\eta_a,\gamma_a,\delta_b,\eta_b,\gamma_b).$ 
Similarly, for the generator the training parameters are 
$\vec{\theta}_{\mathcal{G},\lambda}=(\delta_g,\eta_g,\gamma_g).$ 
After the training is completed and parameters 
$\vec{\theta}_{\mathcal{D},\lambda}$ and 
$\vec{\theta}_{\mathcal{G},\lambda}$ are found we can compute the 
explicit forms of $U,A,B$ (i.e., $\mathcal{D}$) and $\mathcal{G}.$ 
Note that this has to be repeated for each $\lambda$ and that for 
a given $\lambda,$  we can express operator $U$ as
\begin{equation}
U = \cos\frac{\phi}{2} -i(\hat{m}\cdot\vec{\sigma})\sin\frac{\phi}{2}.
\end{equation}
By solving Eq.~(\ref{eq:solve}) we obtain
\begin{equation}
m_x = 0,\quad
m_y = n_z,\quad
m_z = n_z,\quad
\delta = -2 \phi.
\end{equation}
As a summary, the concept of the dedicated minimal (in the sense of 
number controlled unitary gates) SQGEN experiment is given 
in Fig.~\ref{fig:7}b, where the controlled block is implemented as discussed above. 
As already mentioned, it consists of only one 
controlled phase gate and single-qubit rotations. Such experiment could 
be performed on the platform of linear optics~(see, e.g., 
\cite{Lemr2015prl}). Moreover, dedicated circuits for generator output 
of higher dimensions (the dimension of the generator state, which corresponds to the number of qubits) can be constructed in a similar fashion by reducing 
the number of the controlled-unitary gates.

\end{document}